\documentclass[11pt,a4paper]{article}

\usepackage{epsfig}

\usepackage{amsmath}
\usepackage{amssymb}
\usepackage{cite}

\newcommand{\beqs}{\begin{equation*}}
\newcommand{\beq}{\begin{equation}}

\newcommand{\eeqs}{\end{equation*}}
\newcommand{\eeq}{\end{equation}}

\newcommand{\beqas}{\begin{eqnarray*}}
\newcommand{\beqa}{\begin{eqnarray}}

\newcommand{\eeqas}{\end{eqnarray*}}
\newcommand{\eeqa}{\end{eqnarray}}

\newcommand{\eq}[2]{\begin{equation} #1 \label{#2} \end{equation}}

\newcommand{\eps}{\varepsilon}

\newcommand{\om}{\omega}

\newcommand{\la}{\lambda}

\newcommand{\La}{\Lambda}

\newcommand{\blist}{\begin{itemize}}

\newcommand{\elist}{\end{itemize}}

\providecommand{\href}[2]{#2}

\DeclareFontFamily{OT1}{rsfs}{}
\DeclareFontShape{OT1}{rsfs}{m}{n}{ <-7> rsfs5 <7-10> rsfs7 <10->rsfs10}{} 
\DeclareMathAlphabet{\mycal}{OT1}{rsfs}{m}{n}
\newcommand{\scri}{{\mycal I}}

\providecommand{\href}[2]{#2}

\title{HOW TO APPROACH QUANTUM GRAVITY --
BACKGROUND INDEPENDENCE IN 1+1 DIMENSIONS}

\author{
D.~Grumiller\footnote{grumil@hep.itp.tuwien.ac.at} {} and W.~Kummer\footnote{wkummer@tph.tuwien.ac.at}\\Institute f. Theor. Physics\\Vienna University of Technology\\Wiedner Hauptstr. 8-10, A-1040 Vienna, Austria
}

\begin{document}

\maketitle

\begin{abstract}

The application of quantum theory to gravity is beset with
many technical and conceptual problems. After a short tour d'horizon
of recent attempts to master those problems by the introduction of
new approaches, we show that the aim, a background independent quantum
theory of gravity, can be reached in a particular area, 2d dilaton quantum
gravity, {\it without} any assumptions beyond standard
quantum field theory.

\end{abstract}

\section{Introduction}\label{se:1}

It has been realized for some time that a merging
of quantum theory with Einstein's theory of general relativity\footnote{Several reviews on quantum gravity have emerged at the turn of the millennium, cf.\ e.g.\  \cite{Horowitz:1996qd,Carlip:2001wq}.} (GR)
is necessitated by consistency arguments. In {\it Gedankenexperimenten}
the interaction of a classical gravitational wave with a quantum system
inevitably leads to contradictions \cite{Eppley:1977}. Arguments of this type are important because no relevant experimental data are available -- we are very far from the quantum gravity analogue of the Balmer series.

On the other hand, when a quantum theory (QT) of gravity is
developed along usual lines, one is confronted with a fundamental
problem, from which many other (secondary) difficulties can be traced.
The crucial difference to quantum field theory
(QFT) in flat space is the fact that the variables of gravity exhibit
a dual role, they are fields living on a manifold which is determined
by themselves, {}``stage'' and {}``actors'' coincide. But there exist also
numerous other problems: 
the time variable, an object with special properties already
in QT, in GR appears on an equal footing with the space coordinates 
(``problem of time'' which manifests itself in many disguises); the information paradox
\cite{Banks:1995ph}; perturbative non-renormalizability \cite{'tHooft:1974bx} etc.

In section \ref{se:2} we discuss some key-points regarding the definition
of physical observables in QFT and the ensuing ones in quantum gravity
(QGR). Then we critically mention some {}``old'' and {}``new''
approaches to QGR (section \ref{se:3}) from a strictly quantum field theorist's
point of view. Finally we give some highlights on the ``Vienna approach''
to 2d dilaton quantum gravity with matter, including a new result (within that approach) on entropy corrections which is in agreement with the one found in literature (section \ref{se:4}). In that area which contains also
models with physical relevance (e.g.\ spherically reduced gravity)
the application of just the usual concepts of (even nonpertubative!)
QFT lead to very interesting consequences \cite{Grumiller:2002nm} which allow
physical interpretations in terms of {}``solid'' traditional QFT
observables.

\section{Observables}\label{se:2}

\subsection{Cartan variables in GR}

Physical observables in the sense used here are certain functionals
of the field variables which are directly accessible to experimental
measurements.

The metric $g$ in GR can be considered as a {}``derived'' field variable
\begin{equation}
g=e^{a}\otimes e^{b}\eta_{ab},\label{eq:1}\end{equation}
because it is the direct product of the dual basis
one forms\footnote{For details on gravity in the Cartan formulation we refer to the mathematical literature, e.g.\ \cite{nakaharageometry}} $e^{a}=e_{\mu}^{a}\, dx^{\mu}$ contracted with the flat local Lorentz
metric $\eta_{ab}$ which is used to raise and lower ``flat indices'' denoted by Latin letters ($\eta={\rm diag}(1,-1,-1,-1,....),\,\, x^{\mu}=\left\{ x^{0},x^{i}\right\}$).
Local Lorentz invariance leads to the ``covariant derivative'' $D^{a}{}_b=\delta_{b}^{a}d+\omega^{a}{}_b$
with a spin connection 1-form $\omega^{a}{}_b$
as a gauge field. Its antisymmetry $\omega^{ab}=-\omega^{ba}$ implies metricity. Thanks to the Bianchi identities all covariant tensors relevant for constructing actions in even dimensions 
can be expressed in terms of $e^{a}$, the curvature 2-form
$R^{ab}=(D\omega)^{ab}$ and the torsion 2-form $T^{a}=(De)^{a}$.
For nonvanishing torsion the affine connection $\Gamma_{\mu\nu}$$\,^{\rho}=E_{a}^{\rho}\,\left(D_{\mu}e\right)_{\nu}^{a}$,
expressed in terms of components $e_{\mu}^{a}$ and of its inverse
$E_{a}^{\rho}$, besides the usual Christoffel symbols also contains
a contorsion term in $\Gamma_{\left(\mu\nu\right)}\,^{\rho}$, whereas
$\Gamma_{\left[\mu\nu\right]}\,^{\rho}$ are the components of torsion.
Einstein gravity in d=4 dimensions postulates vanishing
torsion $T^{a}=0$ so that $\omega=\omega(e)$. This theory can be
derived from the Hilbert action ($G_{N}$ is Newton's constant; dS space results
for nonvanishing cosmological constant $\Lambda$ from the replacement $R^{ab}\rightarrow R^{ab}-\frac{4}{3}\Lambda e^{a}\wedge e^{b}$)
\begin{equation}
L_{\left(H\right)}=\frac{1}{16\pi G_{n}}\int_{\mathcal{M}_{4}}\, R^{ab}\wedge e^{c}\wedge e^{d}\epsilon_{abcd}+L_{\left({\rm matter}\right)}.\label{eq:2}\end{equation}
Because of the {}``Palatini mystery'', independent
variation of $\delta\omega$ yields $T^{a}=0,$ whereas $\delta e$
produces the Einstein equations.

Instead of working with the metric (\ref{eq:1}) the {}``new''
approaches \cite{Sen:1982qb} are based upon a gauge field related to
$\omega^{ab}$
\begin{equation}
A^{ab}=\frac{1}{2}\left(\omega^{ab}-\frac{\gamma}{2}\epsilon^{ab}{}_{cd}\,\omega^{cd}\right).\label{eq:3}\end{equation}
The Barbero-Immirzi parameter $\gamma$ \cite{Barbero:1995ap} is an arbitrary constant. 
The extension to complex gravity
$\left(\gamma=i\right)$ makes $A^{a}$ a self-adjoint field and transforms
the Einstein theory into the one of an $SU\left(2\right)$ gauge field
\begin{equation}
A_{i}^{\underline{a}}=\epsilon^{0\,\underline{a}}\,_{\underline{b}\,\underline{c}}\, A_{i}^{\,\underline{b}\,\underline{c}},\label{eq:4}\end{equation}
where the index $\underline{a}=1,2,3.$ This formulation
is the basis of loop quantum gravity and spin foam models (see below).

\subsection{Observables in classical GR}

The exploration of the global properties of a certain solution
of (\ref{eq:2}), its singularity structure etc., is only possible
by means of the introduction of an additional test field, most simply
a test particle with action
\begin{eqnarray}
L_{\left({\rm test}\right)} & = & -m_{0}\int\left|ds\right|,\nonumber \\
ds^{2} & = & g_{\mu\nu}\left(x\left(\tau\right)\right)\frac{dx}{d\tau}^{\mu}\frac{\, dx\,^{\nu}}{d\tau},\label{eq:5}\end{eqnarray}
which is another way to incorporate Einstein's old
proposal \cite{Einstein:1916vd} of a {}``net of geodesics''. The
path $x^{\mu}\left(\tau\right)$ is parameterized by the affine parameter
$\tau$ (actually only timelike or lightlike $ds^{2}\geq 0$ describes the paths of
a physical particle).

It is not appreciated always that the global properties of
a manifold are \textit{\large defined} in terms of a specific
device like (\ref{eq:5}). Whereas the usual geodesics derived from
(\ref{eq:5}) depend on $g_{\mu\nu}$ through the Christoffel symbols
only, e.g.\ in the case of torsion also the contorsion may contribute
({}``autoparallels'') in the affine connection; spinning particles
{}``feel'' the gravimagnetic effect etc. As a consequence, when
a field dependent transformation of the gravity variables is performed
(e.g.\ a conformal transformation from a {}``Jordan frame'' to an
{}``Einstein frame'' in Jordan-Brans-Dicke \cite{Fierz:1956} theory)
the action of the device must be transformed in the same way.

\subsection{Observables in QFT}

In flat QFT one starts from a Schr\"odinger
equation, dependent on field operators and, proceeding
through Hamiltonian quantization to the path integral, the experimentally
accessible observables are the elements of the S-matrix, or quantities
expressible by those.\footnote{Note that ordinary quantum mechanics and its Schr\"odinger equation
appear as the nonrelativistic, weak coupling limit of the Bethe-Salpeter
equation of QFT \cite{Salpeter:1951sz}. Useful notions like eigenvalues of Hermitian operators,
collapse of wave functions etc.\ are not basic concepts in this more
general frame (cf.\ footnote 2 in ref.\ \cite{Kummer:2001ip}).}
It should be recalled that the properly defined renormalized
S-matrix element obtains by amputation of external propagators in
the related Green function, multiplication with polarizations and
with a square root of the wave function renormalization constant,
taking the mass-shell limit.

In gauge theories  one
encounters the additional problem of gauge-de\-pen\-dence, i.e.\ the dependence
on some gauge parameter $\beta$ introduced by generic gauge fixing.
Clearly the S-matrix elements must be and indeed are \cite{Kummer:2001ip}
independent of $\beta$. But other objects, in particular matrix-elements
of gauge invariant operators $\mathcal{O}_{A}$, depend on $\beta$.
In addition, under renormalization they mix with operators $\tilde{\mathcal{O}}_{\tilde{A}}$
of the same {}``twist'' (dimension minus spin) which depend on Faddeev-Popov
ghosts \cite{Dixon:1974ss} and are not gauge-invariant:
\begin{eqnarray}
&& \mathcal{O}_{A}^{\left(ren\right)} = Z_{AB}\mathcal{O}_{B}+ Z_{A\tilde{B}}\tilde{\mathcal{O}}_{\tilde{B}}\nonumber \\
&& \tilde{\mathcal{O}}_{\tilde{A}}^{\left(ren\right)} =  \phantom{Z_{AB}\mathcal{O}_{B}+\ } Z_{\tilde{A}\tilde{B}}\tilde{\mathcal{O}}_{\tilde{B}}\label{eq:6}\end{eqnarray}
The contribution of such operators to the S-matrix element
(sic!) of e.g.\ the scaling limit for deep inelastic scattering \cite{Friedman:1991} of leptons on protons \cite{Gross:1973ju} occurs only through the anomalous dimensions 
($\propto\partial Z_{AB}/\partial\Lambda$ for a regularisation
cut-off $\La$). And those objects, also thanks to the triangular form of
(\ref{eq:6}), are gauge-independent!

In flat QFT, as well as in QGR, the (gauge invariant) {}``Wilson
loop''
\begin{equation}
W_{\left(\mathcal{C}\right)}={\rm Tr}\, P\exp{\big(i\oint\limits _{\mathcal{C}}A_{\mu\, dx^{\mu}}\big)},\label{eq:7}\end{equation}
parameterized by a path ordered closed curve $\mathcal{C},$
often is assumed to play an important role. In covariant gauges it
is multiplicatively renormalizable with the renormalization constant
depending on the length of $\mathcal{C}$, the UV cut-off and eventual
cusp-angles in $\mathcal{C}$ \cite{Polyakov:1980ca}. Still the relation
to experimentally observable quantities (should one simply drop the renormalization constant or proceed \cite{Kummer:2001ip} as for an S-matrix?) is unclear. Worse, for lightlike axial gauges $\left(nA\right)=0$ $\left(n^{2}=0\right)$ multiplicative renormalization is not
applicable \cite{Andrasi:1999hu}. Then, only for a matrix element of (\ref{eq:7}) between
{}``on-shell gluons'', this type of renormalization is restored.
Still the renormalization constant is different from covariant gauge,
except for the anomalous dimension derived from it (cf.\ precisely that
feature of operators in deep inelastic scattering).

\section{Approaches to QGR}\label{se:3}

``Old'' QGR worked with a separation of the two aspects
of gravity variables by the decomposition of the metric
\begin{equation}
g_{\mu\nu}=g_{\mu\nu}^{\left(0\right)}+h_{\mu\nu},\label{eq:8}
\end{equation}
which consists of a (fixed) classical background $g_{\mu\nu}^{\left(0\right)}$
({}``stage'') with small quantum fluctuations $h_{\mu\nu}$ ({}``actors'').
The {}``observable'' (to be tested by a classical device) would
be the effective matrix $g_{\mu\nu}^{\left({\rm eff}\right)}=g_{\mu\nu}^{\left(0\right)}+<h_{\mu\nu}>.$
Starting computations from the action (\ref{eq:2}) one finds that an ever increasing number of counter-terms is necessary. They are different from the terms in
the Lagrangian $\mathcal{L}=\sqrt{-g}\, R/\left(16\pi G_{N}\right)$
in (\ref{eq:2}). This is the reason why QGR is called (perturbatively) {}``nonrenormalizable''
\cite{'tHooft:1974bx}. Still, at energies E $\ll\left(G_{N}\right)^{-1/2}$,
i.e.\ much below the Planck mass scale $m_{\rm Pl}\propto\left(G_{N}\right)^{-1/2}$,
such calculations can be meaningful in the sense of an {}``effective
low energy field theory'' \cite{Donoghue:1994dn}, irrespective of the fact that (perhaps
by embedding gravity into string theory) by inclusion of further fields
at higher energy scales (Planck scale), QGR may become renormalizable.
Of course, such an approach even when it is modified by iterative
inclusion of $<h_{\mu\nu}>$ into $g_{\mu\nu}^{\left(0\right)}$ etc.\
-- which is technically quite hopeless -- completely misses inherent
background independent effects, i.e.\ effects when $g_{\mu\nu}^{\left(0\right)}=0$.

One could think also of applying nonperturbative methods developed
in numerical lattice calculations for QCD. However, there are problems
to define the Euclidean path integral for that, because the Euclidean
action is not bounded from below (as it is the case in QCD) \cite{Gibbons:1994cg}.

The quantization of gravity which -- at least in principle
-- avoids background dependence is based upon the ADM approach to the
Dirac quantization of the Hamiltonian \cite{Arnowitt:1962}. Space-time is foliated
by a sequence of three dimensional space-like manifolds $\sum_{3}$
upon which the variables $g_{ij}=q_{ij}$ and associated canonical
momenta $\pi_{ij}$ live. The constraints associated to the further
variables lapse $\left(N_{0}\right)$ and shift $\left(N_{i}\right)$
in the Hamiltonian density
\begin{equation}
\mathcal{H}=N_{0}\, H^{0}\left(q,\pi\right)+N_{i}\, H^{i}\left(q,\pi\right)\label{eq:9}\end{equation}
are primary ones. The Poisson brackets of the secondary constraints
$H^{\mu}$ closes. $H^{i}$ generates diffeomorphisms on $\sum_{3}$.
In the quantum version of (\ref{eq:9}) the solutions of the Wheeler-deWitt
equation involving the Hamiltonian constraint
\begin{equation}
\int\limits _{\sum_{3}}\, H^{0}\,\left(q,\,\frac{\delta}{i\delta q}\right)\mid\psi>=0\label{eq:10}\end{equation}
 formally would correspond to a nonperturbative QGR. Apart
from the fact that it is extremely difficult to find a general solution
to (\ref{eq:10}) there are several basic problems with a quantum
theory based upon that equation (e.g.\ no Hilbert space $\mid\psi>$
can be constructed, no preferred time foliation exists with ensuing inequivalent
quantum evolutions \cite{Torre:1998eq}, problems with usual {}``quantum
causality'' exist, the {}``axiom'' that fields should commute at
space like distances does not hold etc.). A restriction to a finite
number of degrees of freedom ({}``mini superspace'') \cite{Dewitt:1967yk} or infinite number of degrees of freedom (but still less than the original theory -- so-called ``midi superspace'') \cite{Kuchar:1971xm}
has been found to miss essential features.

As all physical states $\mid\psi>$ must be annihilated by
the constraints $H^{\mu}$, a naive Schr\"odinger equation involving
the Hamiltonian constraint $H^{0}$,
\begin{equation}
i\hbar\frac{\partial\mid\psi>}{\partial t}=H^{0}\mid\psi>=0,\label{eq:11}
\end{equation}
cannot contain a time variable ({}``problem of time'').
A kind of Schr\"odinger equation can be produced by the definition of
a {}``time-function'' $T\left(q,\pi,x\right)$, at the price of
an even more complicated formalism \cite{Bergmann:1962} with quite ambiguous
results -- and the problem, how to connect those with {}``genuine''
observables. All these problems are aggravated, when one tries to
first eliminate constraints by solving them explicitly before quantization.
In this way, clearly part of the quantum fluctuations are eliminated
from the start. As a consequence different quantum theories, constructed
in this way, are not equivalent.

The {}``new'' gravities (loop quantum gravity, spin foam models) reformulate the quantum theory of space-time
by the introduction of novel variables, based upon the concept of
Wilson loops (\ref{eq:7}) applied to the gauge-field (\ref{eq:4}).
The operator
\begin{equation}
U\left(s_{1},s_{2}\right)={\rm Tr}\, P\, \exp{\left( i\int\limits _{s_{1}}^{s_{2}}ds\frac{dx^{i}}{ds}\, A_{i}\right)} \label{eq:12}
\end{equation}
defines a holonomy. It is generalized by inserting further
invariant operators at intermediate points between $s_{1}$ and $s_{2}$
. From such holonomies a spin network can be created which represents spacetime
(in the path integral it is dubbed ``spin foam'').

These approaches claim several successes \cite{Carlip:2001wq}. Introducing
as a basis diffeomorphism equivalence classes of {}``labeled graphs''
a finite Hilbert space can be constructed and some solutions of the
Wheeler-deWitt equation (\ref{eq:10}) have been obtained. The methods
introduce a {}``natural'' coarse graining of space-time which implies
a $UV$ cutoff. {}``Small'' gravity around certain states leads
in those cases to corresponding linearized Einstein gravity.

However, despite of very active research in this field a number
of very serious open questions persists: The Hamiltonian constructed
from spin networks does not lead to massless excitations (gravitons)
in the classical limit. The Barbero-Immirzi parameter $\gamma$ has to be fixed by
the requirement of a ``correct'' Bekenstein-Hawking entropy for
the Black Hole. The most severe problem, however, is the one of observables.
By some researchers in this field it has been claimed that by {}``proper
gauge fixing'' (!) area and volume can be obtained as quantized {}``observables'',
which is a contradiction in itself from the point of view of QFT.
We must emphasize too that also in an inherently $UV$ regularized
theory (finite) renormalization remains an issue to be dealt with
properly. Also the fate of S-matrix elements, which play such a central
role as the proper observables in QFT, is completely unclear in these
setups.

Embedding QGR into (super-)string theory \cite{Polchinski:1998rq} does not remove the
key problems related to the dual role of the metric. Gravity may well be a string excitation
in a string/brane world of 10-11 dimensions, possibly a finite
theory of everything. Nevertheless, at low energies Einstein gravity (eventually
plus an antisymmetric $B$-field) remains the theory for which computations
must be performed.\footnote{It should be noted that the now widely confirmed astronomical
observations of a positive cosmological constant \cite{Riess:2001gk}
(if it is a constant and not a {}``quintessence'' field in a theory
of type \cite{Fierz:1956}) precludes immediate application of supersymmetry
(supergravity) in string theory, because only AdS space
is compatible with supergravity \cite{VanNieuwenhuizen:1981ae}.}
Unfortunately, the proper choice (let alone the derivation)
of a string vacuum in our d=4 space-time is an unsolved problem.


Many other approaches exist, including noncommutative geometry, twis\-tors, causal sets, 3d approaches, dynamical triangulations, Regge calculus etc., each of which has certain attractive features and difficulties (cf.\ e.g.\ \cite{Carlip:2001wq} and refs.\ therein).

To us all these {}``new'' approaches appear as -- very ingenious
-- attempts to bypass the technical problems of directly applying standard
QFT to gravity -- without a comprehensive solution of the main problems of QGR being in sight. Thus the
main points of a {}``minimal'' QFT for gravity should be based upon
``proven concepts'' of QFT with a point of departure characterizing QGR
as follows:
\begin{itemize}
\item[(a)] QGR is an {}``effective'' low energy theory and therefore need not
be renormalizable to all orders.
\item[(b)] QGR is based upon classical Einstein (-dS) gravity with usual
variables (metric or Cartan variables).
\item[(c)] At least the quantization of geometry must be performed in a background independent (nonperturbative) way.
\item[(d)] Absolutely {}``safe'' quantum observables are only the S-matrix
elements of QFT ${<f\mid S\mid i>}$, where initial state ${\mid i>}$
and final state ${<f\mid}$ are defined only when those states are
realized as Fock states of particles in a (at least approximate) flat
space environment. In certain cases it is permissible to employ a
semi-classical approach: expectation values of quantum corrections
may be added to classical geometric variables, and a classical computation
is then based on the effective variables, obtained in this way.
\end{itemize}
Clearly item (d) by construction
excludes any application to quantum cosmology, where $\mid i>$ would
be the (probably nonexistent) infinite past before the Big Bang.

Obviously the most difficult issue is (c). We describe in the following
section how gravity models in d=2 (e.g.\ spherically reduced gravity)
permit a solution of just that crucial point, leading to novel results.

\section{``Minimal'' QGR in 1+1 dimensions}\label{se:4}

\subsection{Classical theory: first order formulation}

In the 1990s the interest in dilaton gravity in d=2 was rekindled by results from string theory \cite{Mandal:1991tz}, but it existed as a field on its own more or less since the 1980s \cite{Barbashov:1979}. For a review on dilaton gravity ref.\ \cite{Grumiller:2002nm} may be consulted. For sake of self-containment the study of dilaton gravity will be motivated briefly from a purely geometrical point of view.

The notation of ref.\ \cite{Grumiller:2002nm} is used: $e^a$ is the
zweibein one-form, $\epsilon = e^+\wedge e^-$ is the volume two-form. The one-form
$\omega$ represents the  spin-connection $\om^a{}_b=\eps^a{}_b\om$
with  the totally antisymmetric Levi-Civit{\'a} symbol $\eps_{ab}$ ($\eps_{01}=+1$). With the
flat metric $\eta_{ab}$ in light-cone coordinates
($\eta_{+-}=1=\eta_{-+}$, $\eta_{++}=0=\eta_{--}$) the torsion 2-form reads
$T^\pm=(d\pm\omega)\wedge e^\pm$. The curvature 2-form $R^a{}_b$ can be presented by the 2-form $R$ defined by 
$R^a{}_b=\eps^a{}_b R$, $R=d\wedge\om$.
Signs and factors of the Hodge-$\ast$ operation are defined by $\ast\epsilon=1$. 

Since the Hilbert action $\int_{\mathcal{M}_2}  R\propto(1-g)$ yields just the Euler number for a surface with genus $g$ one has to generalize it appropriately. The simplest idea is to introduce a Lagrange multiplier for curvature, $X$, also known as ``dilaton field'', and an arbitrary potential thereof, $V(X)$, in the action $\int_{\mathcal{M}_2}  \left(XR+\epsilon V(X)\right)$. In particular, for $V\propto X$ the Jackiw-Teitelboim model emerges \cite{Barbashov:1979}. Having introduced curvature it is natural to consider torsion as well. By analogy the first order gravity action \cite{Ikeda:1993aj}
\eq{
L^{(1)}=\int_{\mathcal{M}_2}  \left(X_aT^a+XR+\epsilon\mathcal{V} (X^aX_a,X)\right)
}{eq:FOG}
can be motivated where $X_a$ are the Lagrange multipliers for torsion. It encompasses essentially all known dilaton theories in 2d, also known as Generalized Dilaton Theories (GDT). Spherically reduced gravity (SRG) from d=4 corresponds to $\mathcal{V} = -X^+X^-/(2X)-{\rm const}$.

Without matter there are no physical propagating degrees of freedom, which is advantageous mathematically but not very attractive from a physical point of view. Thus, in order to describe scattering processes matter has to be added. The simplest way is to consider a massless Klein-Gordon field $\phi$,
\eq{
L^{(m)} = \frac{1}{2} \; \int_{\mathcal{M}_2}\;
F(X)\, d \phi  \wedge \ast d \phi\,, 
}{eq:matter}
with a coupling function $F(X)$ depending on the dilaton (for dimensionally reduced theories typically $F\propto X$ holds). 

\subsection{Quantum theory: Virtual Black Holes}

It turned out that even in the presence of matter an exact path integration 
of all geometric quantities is possible for all GDTs, proceeding along well 
established paths of QFT\footnote{We mention just a few technical details: no ordering ambiguities arise, the (nilpotent) BRST charge is essentially the same as for Yang-Mills theory (despite of the appearance of nonlinearities in the algebra of the first class secondary constraints), the gauge fixing fermion is chosen such that ``temporal'' gauge is obtained, the Faddeev-Popov determinant cancels after integrating out the ``unphysical'' sector, and ``natural'' boundary conditions cannot be imposed on the fields, so one has to be careful with the proper treatment of the boundary.} \cite{Kummer:1992rt}. 

The effective theory obtained in 
this way solely depends on the matter fields in which it is nonlocal and 
non-polynomial. Already at the level of the (nonlocal) vertices of matter 
fields, to be used in a systematic perturbative expansion in terms of Newton's 
constant, a highly nontrivial and physically intriguing phenomenon can be 
observed, namely the so-called  ``virtual black hole'' (VBH). This notion 
originally has been introduced by S. Hawking \cite{Hawking:1996ag}, but in our 
recent approach the VBH for SRG emerges naturally
in Minkowski signature space-time, without the necessity of additional {\em ad 
hoc} assumptions. 

For non-minimally coupled scalars the 
lowest order S-matrix indeed exhibited interesting
features: forward scattering poles, monomial scaling with energy, CPT 
invariance, and pseudo-self-similarity in its kinematic sector 
\cite{Grumiller:2000ah}.

\begin{figure}
\center
\includegraphics[width=80pt]{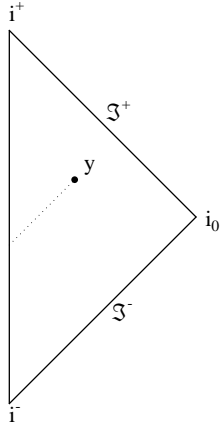}
\caption{CP diagram of the VBH}
\label{fig:cp}
\end{figure}
It was possible to reconstruct geometry self-consistently from a 
(perturbative or, if available, exact) solution of the effective theory. For
the simplest case of four-point tree-graph scattering the corresponding 
Carter-Penrose (CP) diagram is presented in Fig. \ref{fig:cp}. It is non-local 
in the sense that it depends not only on one set of coordinates but on two. 
This was a consequence of integrating out geometry non-perturbatively. For 
each choice of $y$ (one of the two sets of coordinates) it is possible to draw 
an ordinary CP-diagram. The non-trivial part of our effective geometry (i.e.\
the VBH) is concentrated on the light-like cut. For SRG the ensuing 
line-element has Sachs-Bondi form
\begin{equation}
(ds)^2 = 2 dr du + \left(1 - \frac{2m(u,r)}{r} - a(u,r) r + d(u,r)\right) (du)^2\,,
\end{equation}
with $m$, $a$ and $d$ being localized\footnote{The localization of ``mass'' 
and ``Rindler acceleration'' on a light-like cut  
is not an artifact of an accidental gauge choice, but has a physical 
interpretation in terms of the Ricci-scalar. Certain parallels to Hawking's
Euclidean VBHs can be observed, but also essential differences. The main one 
is our Minkowski signature which we deem to be a positive feature.} 
on the cut $u=u_0$ with compact support
$r<r_0$. These quantities depend on the second set of coordinates $u_0$, $r_0$.

One should not take the effective geometry at face value -- this would be like 
over-interpreting the role of virtual particles in a loop diagram. It is a 
nonlocal entity and one still has to ``sum'' (read: integrate) over all 
possible geometries of this type in order to obtain the nonlocal vertices and 
the scattering amplitude. Nonetheless, the simplicity of this geometry and the 
fact that all possible configurations are summed over are nice features of 
this picture. Moreover, all VBH geometries coincide asymptotically and differ only very little from each other in the asymptotic region. This observation allows for the following interpretation: the boundaries of the diagram, $\scri^\pm$ and $i^0$, behave in a classical way\footnote{Clearly the imposed boundary conditions play a crucial role in this context. They produce effectively a fixed background, but only at the boundary.} (thus enabling one to construct an ordinary Fock space like in fixed background QFT), but the more one zooms into the geometry the less classical it becomes. The situation seems to be quite contrary to Kucha\v{r}'s proposal of geometrodynamics\footnote{This approach considers only the matterless case and thus a full comparison to our results is not possible.} of BHs: while we have fixed boundary conditions for the target space coordinates (and hence a fixed ADM mass) but a ``smeared geometry'' (in the sense that a continuous spectrum of asymptotically equivalent VBHs contributes to the S-matrix), Kucha\v{r} encountered a ``smeared mass'' (obeying a Schr\"odinger equation) but an otherwise fixed geometry \cite{Kuchar:1994zk}.

Qualitatively it is clear what has to be done in order to obtain the 
S-matrix\footnote{The idea that BHs must be considered in the S-matrix 
together with 
elementary matter fields has been put forward some time ago 
\cite{'tHooft:1996tq}. The approach \cite{Grumiller:2000ah} 
reviewed here, for the first time allowed to derive (rather than to 
conjecture) the appearance of the BH states in the quantum scattering matrix 
of gravity.}: 
Take all possible VBHs of Fig.\ \ref{fig:cp} and sum them coherently with proper
weight factors and suitably attached external legs of scalar fields. This had
been done quantitatively in a straightforward but rather lengthy calculation
for gravitational scattering of s-waves in the framework of SRG, the result of 
which yielded the lowest order tree-graph S-matrix for ingoing modes with 
momenta $q,q'$ and outgoing ones $k,k'$,
\begin{equation}
T(q, q'; k, k') = -\frac{i\kappa\delta\left(k+k'-q-q'\right)}{2(4\pi)^4 
|kk'qq'|^{3/2}} E^3 \tilde{T}\,,
\end{equation}
with the total energy $E=q+q'$, $\kappa=8\pi G_N$, 
\begin{eqnarray}
\tilde{T} (q, q'; k, k') := \frac{1}{E^3}{\Bigg [}\Pi \ln{\frac{\Pi^2}{E^6}}
+ \frac{1} {\Pi} \sum_{p \in \left\{k,k',q,q'\right\}}  
\nonumber \\
p^2 \ln{\frac{p^2}{E^2}} {\Bigg (}3 kk'qq'-\frac{1}{2}
\sum_{r\neq p} \sum_{s \neq r,p}\left(r^2s^2\right){\Bigg )} {\Bigg ]}\,,
\end{eqnarray}
and the momentum transfer function $\Pi = (k+k')(k-q)(k'-q)$. The interesting 
part of the scattering amplitude is encoded in the scale independent factor 
$\tilde{T}$. The forward scattering poles occurring for $\Pi=0$ should be noted.

It is possible to generalize the VBH phenomenon to arbitrary GDTs with
matter as well as most of its properties (for instance, the CP-diagram, CPT 
invariance and the role played in the S-matrix) \cite{Grumiller:2002dm}.

\subsection{New results and outlook}

Recently quantum corrections to the specific heat of the dilaton BH have been calculated by applying the quantization method discussed above \cite{Grumiller:2003mc}. The result is $C_s:=dM/dT=96\pi^2M^2/\la^2$, where $\la$ is the scale parameter of the theory. Thus, in that particular case quantum corrections lead to a stabilization of the system. The mass of the BH is found to be decreasing according to
\eq{
M(u)\approx M_0-\frac{\pi}{6} (T_H^0)^2(u-u_0)-\frac{\la}{24\pi}\ln{\frac{M(u)}{M_0}} + {\mathcal O}\left(\frac{\la}{M(u)}\right)\,.
}{referee:2}
The first term is the ADM mass, the second term corresponds to a linear decrease due to the (in leading order) constant Hawking flux and the third term provides the first nontrivial correction.

Applying simple thermodynamical methods\footnote{The review \cite{Wald:1999vt} may be consulted in this context.} ($dS=C_sdT/T$) and exploiting the quantum corrected mass/temperature relation $T/T_0=1-\la/(48\pi M)$ it is possible to calculate also entropy corrections:
\eq{
S = S_0 - \frac{1}{24} \ln{S_0} + {\mathcal O}(1)\,,\quad S_0:=\frac{2\pi M}{\la}=\left. 2\pi X\right|_{\rm horizon}
}{eq:entropy}
The logarithmic behavior is in qualitative agreement with the one found in the literature by various methods \cite{Mann:1998hm}; the prefactor $1/24$ coincides with \cite{Zaslavskii:1996dg}.  

An extension of the results obtained in the first order formulation to dilaton supergravity is straightforward in principle but somewhat tedious in detail. It permitted, among other results, to obtain for the first time a full solution of dilaton supergravity \cite{Bergamin:2003am}.

All these exciting applications indicate that the strict application of standard QFT concepts to gravity (at least in d=2 or in models dimensionally reduced to d=2) shows great promise.

\section*{Acknowledgement}
 
This work has been supported by project P-14650-TPH of the Austrian Science Foundation (FWF) and by EURESCO. We are grateful to the organizers of the workshop ``What comes beyond the Standard Model?'' for an enjoyable meeting in Slovenia. We thank M.\ Bojowald and O.\ Zaslavskii for useful discussions at the Erwin Schr\"odinger Institut in Vienna, L.\ Bergamin for collaboration on 2d dilaton supergravity and D.\ Vassilevich for a long time collaboration on 2d dilaton gravity.


\begin{thebibliography}{99}

\bibitem{Horowitz:1996qd}
G.~T. Horowitz,
\newblock (1996), gr-qc/9604051;
C.~Rovelli,
\newblock Living Rev. Rel. {\bf 1}, 1 (1998), gr-qc/9710008;
\newblock Notes for a brief history of quantum gravity,
\newblock in {\em 9th {M}arcel {G}rossmann Meeting On Recent Developments In
  Theoretical And Experimental General Relativity, Gravitation And Relativistic
  Field Theories}, 2000, arXiv:gr-qc/0006061.

\bibitem{Carlip:2001wq}
S.~Carlip,
\newblock Rept. Prog. Phys. {\bf 64}, 885 (2001), arXiv:gr-qc/0108040;
L.~Smolin,
\newblock (2003), hep-th/0303185.

\bibitem{Eppley:1977}
K.~Eppley and E.~Hannah,
\newblock Found. Phys. {\bf 7}, 51 (1977).

\bibitem{Banks:1995ph}
T.~Banks,
\newblock Nucl. Phys. Proc. Suppl. {\bf 41}, 21 (1995), hep-th/9412131.

\bibitem{'tHooft:1974bx}
G.~'t~Hooft and M.~J.~G. Veltman,
\newblock Annales Poincare Phys. Theor. {\bf A20}, 69 (1974).

\bibitem{Grumiller:2002nm}
D.~Grumiller, W.~Kummer, and D.~V. Vassilevich,
\newblock Phys. Rept. {\bf 369}, 327 (2002), hep-th/0204253.

\bibitem{nakaharageometry}
M.~Nakahara,
\newblock {\em {Geometry, Topology and Physics}} (IOP Publishing, Bristol,
  1990).

\bibitem{Sen:1982qb}
A.~Sen,
\newblock Phys. Lett. {\bf B119}, 89 (1982);
A.~Ashtekar,
\newblock Phys. Rev. Lett. {\bf 57}, 2244 (1986);
\newblock Phys. Rev. {\bf D36}, 1587 (1987);
C.~Rovelli,
\newblock Class. Quant. Grav. {\bf 8}, 1613 (1991).

\bibitem{Barbero:1995ap}
J.~F. Barbero,
\newblock Phys. Rev. {\bf D51}, 5507 (1995), gr-qc/9410014;
G.~Immirzi,
\newblock Class. Quant. Grav. {\bf 14}, L177 (1997), gr-qc/9612030;
\newblock Nucl. Phys. Proc. Suppl. {\bf 57}, 65 (1997), gr-qc/9701052.

\bibitem{Einstein:1916vd}
A.~Einstein,
\newblock Annalen Phys. {\bf 49}, 769 (1916).

\bibitem{Fierz:1956}
M.~Fierz,
\newblock Helv. Phys. Acta {\bf 29}, 128 (1956);
P.~Jordan,
\newblock Z. Phys. {\bf 157}, 112 (1959);
C.~Brans and R.~H. Dicke,
\newblock Phys. Rev. {\bf 124}, 925 (1961).

\bibitem{Salpeter:1951sz}
E.~E. Salpeter and H.~A. Bethe,
\newblock Phys. Rev. {\bf 84}, 1232 (1951);
N.~Nakanishi,
\newblock Prog. Theor. Phys. Suppl. {\bf 43}, 1 (1969);
R.~Barbieri and E.~Remiddi,
\newblock Nucl. Phys. {\bf B141}, 413 (1978);
W.~E. Caswell and G.~P. Lepage,
\newblock Phys. Rev. {\bf A18}, 810 (1978).

\bibitem{Kummer:2001ip}
W.~Kummer,
\newblock Eur. Phys. J. {\bf C21}, 175 (2001), hep-th/0104123.

\bibitem{Dixon:1974ss}
J.~A. Dixon and J.~C. Taylor,
\newblock Nucl. Phys. {\bf B78}, 552 (1974);
J.~A. Dixon,
\newblock Nucl. Phys. {\bf B99}, 420 (1975);
H.~Kluberg-Stern and J.~B. Zuber,
\newblock Phys. Rev. {\bf D12}, 467 (1975);
W.~Konetschny and W.~Kummer,
\newblock Nucl. Phys. {\bf B124}, 145 (1977);
S.~D. Joglekar and B.~W. Lee,
\newblock Ann. Phys. {\bf 97}, 160 (1976).

\bibitem{Friedman:1991}
J.~I. Friedman, H.~W. Kendall, and R.~E. Taylor,
\newblock Rev. Mod. Phys. {\bf 63}, 573 (1991).

\bibitem{Gross:1973ju}
D.~J. Gross and F.~Wilczek,
\newblock Phys. Rev. {\bf D8}, 3633 (1973);
\newblock Phys. Rev. {\bf D9}, 980 (1974).

\bibitem{Polyakov:1980ca}
A.~M. Polyakov,
\newblock Nucl. Phys. {\bf B164}, 171 (1980);
V.~S. Dotsenko and S.~N. Vergeles,
\newblock Nucl. Phys. {\bf B169}, 527 (1980);
R.~A. Brandt, F.~Neri, and M.-a. Sato,
\newblock Phys. Rev. {\bf D24}, 879 (1981);
J.~Frenkel and J.~C. Taylor,
\newblock Nucl. Phys. {\bf B246}, 231 (1984).

\bibitem{Andrasi:1999hu}
A.~Andrasi,
\newblock Eur. Phys. J. {\bf C18}, 601 (2001), hep-th/9912138.

\bibitem{Donoghue:1994dn}
J.~F. Donoghue,
\newblock Phys. Rev. {\bf D50}, 3874 (1994), gr-qc/9405057.

\bibitem{Gibbons:1994cg}
G.~W. Gibbons and S.~W. Hawking, editors,
\newblock {\em Euclidean quantum gravity} (Singapore: World Scientific, 1993).

\bibitem{Arnowitt:1962}
R.~Arnowitt, S.~Deser, and C.~W. Misner,
\newblock in {\em {Gravitation: An Introduction to Current Research}}, edited
  by L.~Witten, Wiley, New York, 1962.

\bibitem{Torre:1998eq}
C.~G. Torre and M.~Varadarajan,
\newblock Class. Quant. Grav. {\bf 16}, 2651 (1999), hep-th/9811222.

\bibitem{Dewitt:1967yk}
B.~S. DeWitt,
\newblock Phys. Rev. {\bf 160}, 1113 (1967);
C.~W. Misner,
\newblock Phys. Rev. {\bf 186}, 1319 (1969).

\bibitem{Kuchar:1971xm}
K.~Kucha{\v{r}},
\newblock Phys. Rev. {\bf D4}, 955 (1971).

\bibitem{Bergmann:1962}
P.~Bergmann and A.~Komar,
\newblock in {\em Recent developments in general relativity}, Pergamon, New
  York, 1962;
C.~Teitelboim,
\newblock Phys. Lett. {\bf B56}, 376 (1975);
K.~Kucha\v{r},
\newblock in {\em Quantum gravity 2: an {O}xford symposium}, edited by
  C.~Isham, R.~Penrose, and D.~Sciama, Clarendon Press, Oxford, 1975.

\bibitem{Polchinski:1998rq}
J.~Polchinski,
\newblock {\em String theory} (Cambridge University Press, 1998),
\newblock Vol. 1: {A}n Introduction to the Bosonic String;
\newblock Vol. 2: {S}uperstring Theory and Beyond.

\bibitem{Riess:2001gk}
A.~G. Riess {\em et~al.},
\newblock Astrophys. J. {\bf 560}, 49 (2001), astro-ph/0104455;
A.~G. {Riess},
\newblock {The Publications of the Astronomical Society of the Pacific} {\bf
  112}, 1284 (2000);
Supernova Search Team, A.~G. Riess {\em et~al.},
\newblock Astron. J. {\bf 116}, 1009 (1998), astro-ph/9805201;
Supernova Cosmology Project, S.~Perlmutter {\em et~al.},
\newblock Astrophys. J. {\bf 517}, 565 (1999), arXiv:astro-ph/9812133.

\bibitem{VanNieuwenhuizen:1981ae}
P.~Van~Nieuwenhuizen,
\newblock Phys. Rept. {\bf 68}, 189 (1981).

\bibitem{Mandal:1991tz}
G.~Mandal, A.~M. Sengupta, and S.~R. Wadia,
\newblock Mod. Phys. Lett. {\bf A6}, 1685 (1991);
S.~Elitzur, A.~Forge, and E.~Rabinovici,
\newblock Nucl. Phys. {\bf B359}, 581 (1991);
E.~Witten,
\newblock Phys. Rev. {\bf D44}, 314 (1991);
R.~Dijkgraaf, H.~Verlinde, and E.~Verlinde,
\newblock Nucl. Phys. {\bf B371}, 269 (1992);
C.~G. Callan, Jr., S.~B. Giddings, J.~A. Harvey, and A.~Strominger,
\newblock Phys. Rev. {\bf D45}, 1005 (1992), hep-th/9111056.

\bibitem{Barbashov:1979}
B.~M. Barbashov, V.~V. Nesterenko, and A.~M. Chervyakov,
\newblock Teor. Mat. Fiz. {\bf 40}, 15 (1979);
\newblock J. Phys. {\bf A13}, 301 (1979);
\newblock Theor. Math. Phys. {\bf 40}, 572 (1979);
E.~D'Hoker and R.~Jackiw,
\newblock Phys. Rev. {\bf D26}, 3517 (1982);
C.~Teitelboim,
\newblock Phys. Lett. {\bf B126}, 41 (1983);
E.~D'Hoker, D.~Freedman, and R.~Jackiw,
\newblock Phys. Rev. {\bf D28}, 2583 (1983);
E.~D'Hoker and R.~Jackiw,
\newblock Phys. Rev. Lett. {\bf 50}, 1719 (1983);
R.~Jackiw,
\newblock (1995), hep-th/9501016.

\bibitem{Ikeda:1993aj}
N.~Ikeda and K.~I. Izawa, 
\newblock Prog. Theor. Phys. {\bf 90}, 237--246 (1993), hep-th/9304012;
N.~Ikeda, Ann. Phys. {\bf 235} 435--464 (1994), hep-th/9312059;
P.~Schaller and T.~Strobl, Mod. Phys. Lett. {\bf A9} 3129--3136 (1994), hep-th/9405110.

\bibitem{Kummer:1992rt}
W.~Kummer and D.~J. Schwarz,
\newblock Nucl. Phys. {\bf B382}, 171 (1992);
F.~Haider and W.~Kummer,
\newblock Int. J. Mod. Phys. {\bf A9}, 207 (1994);
W.~Kummer, H.~Liebl, and D.~V. Vassilevich,
\newblock Nucl. Phys. {\bf B493}, 491 (1997), gr-qc/9612012;
\newblock Nucl. Phys. {\bf B544}, 403 (1999), hep-th/9809168;
D.~Grumiller,
\newblock {\em Quantum dilaton gravity in two dimensions with matter},
\newblock PhD thesis, {T}echnische {U}niversit{\"a}t {W}ien, 2001,
  gr-qc/0105078.

\bibitem{Hawking:1996ag}
S.~W. Hawking,
\newblock Phys. Rev. {\bf D53}, 3099 (1996), hep-th/9510029.

\bibitem{Grumiller:2000ah}
D.~Grumiller, W.~Kummer, and D.~V. Vassilevich,
\newblock Nucl. Phys. {\bf B580}, 438 (2000), gr-qc/0001038;
P.~Fischer, D.~Grumiller, W.~Kummer, and D.~V. Vassilevich,
\newblock Phys. Lett. {\bf B521}, 357 (2001), gr-qc/0105034,
\newblock Erratum ibid. {\bf B532} (2002) 373;
D.~Grumiller,
\newblock Int. J. Mod. Phys. {\bf A 17}, 989-992 (2001), hep-th/0111138;
\newblock Class. Quant. Grav. {\bf 19}, 997 (2002), gr-qc/0111097.

\bibitem{Kuchar:1994zk}
K.~V. Kucha{\v{r}},
\newblock Phys. Rev. {\bf D50}, 3961 (1994), arXiv:gr-qc/9403003.

\bibitem{'tHooft:1996tq}
G.~'t~Hooft,
\newblock Int. J. Mod. Phys. {\bf A11}, 4623 (1996), gr-qc/9607022.

\bibitem{Grumiller:2002dm}
D.~Grumiller, W.~Kummer, and D.~V. Vassilevich,
\newblock European Phys. J. {\bf C30}, 135 (2003), hep-th/0208052.

\bibitem{Grumiller:2003mc}
D.~Grumiller, W.~Kummer, and D.~V. Vassilevich,
\newblock JHEP {\bf 07}, 009 (2003), hep-th/0305036.

\bibitem{Wald:1999vt}
R.~M. Wald,
\newblock Living Rev. Rel. {\bf 4}, 6 (2001), gr-qc/9912119.

\bibitem{Mann:1998hm}
R.~B. Mann and S.~N. Solodukhin,
\newblock Nucl. Phys. {\bf B523}, 293 (1998), hep-th/9709064;
R.~K. Kaul and P.~Majumdar,
\newblock Phys. Rev. Lett. {\bf 84}, 5255 (2000), gr-qc/0002040;
S.~Carlip,
\newblock Class. Quant. Grav. {\bf 17}, 4175 (2000), gr-qc/0005017.

\bibitem{Zaslavskii:1996dg}
T.~M.~Fiola, J.~Preskill, A.~Strominger and S.~P.~Trivedi,
\newblock Phys.\ Rev.\ D {\bf 50}, 3987 (1994), hep-th/9403137;
R.~C.~Myers,
\newblock Phys.\ Rev.\ D {\bf 50}, 6412 (1994), hep-th/9405162;
J.~D.~Hayward,
\newblock Phys.\ Rev.\ D {\bf 52}, 2239 (1995), gr-qc/9412065;
O.~B. Zaslavskii,
\newblock Phys. Lett. {\bf B375}, 43-46 (1996).

\bibitem{Bergamin:2003am}
L.~Bergamin and W.~Kummer,
\newblock (2003), hep-th/0306217;
L.~Bergamin, D.~Grumiller, and W.~Kummer,
\newblock (2003), hep-th/0310006.

\end{thebibliography}

\end{document}